\documentclass[10pt]{amsart}

\usepackage{amsmath,amssymb,dsfont}

\usepackage{texdraw} 

\newfont{\bbd}{msbm10 scaled\magstep1}

\def\id{\hbox{{1}\kern-.25em\hbox{\rm l}}}
\def\one#1{#1^{\raise5pt\hbox{$\scriptstyle\!\!\!\!1$}}\,{}}
\def\two#1{#1^{\raise5pt\hbox{$\scriptstyle\!\!\!\!2$}}\,{}}
\def\half{\frac{1}{2}}

\def\comment#1{}

\def\?{(?)\marginpar{|?}}

\def\beq{\begin{equation}}
\def\eeq{\end{equation}}
\def\bea{\begin{eqnarray}}
\def\eea{\end{eqnarray}}
\def\bmat{\left(\begin{array}}
\def\emat{\end{array}\right)}

\newtheorem{theorem}{Theorem}

\makeatletter
\expandafter\let\expandafter
\reset@font\csname reset@font\endcsname
\def\subeqnarray{\arraycolsep1pt
    \def\@eqnnum\stepcounter##1{\stepcounter{subequation}%
        {\reset@font\rm(\theequation\alph{subequation})}}
\jot5mm     \eqnarray}

\makeatother

\newcounter{appendix}
\newcommand{\Ss}{{\mathsf S}}

\newcommand{\RR}{{\mathbb R}}

\def\al{\alpha}
\def\ep{\varepsilon}
\def\l{\lambda}

\def\back{\!\!\!\!\!\!}

\def\mat2#1#2#3#4{{\left(\begin{array}{cc}#1 & #2\\ #3 & #4
      \end{array}\right)}}

\def\mats2#1#2#3#4{{\left(\begin{array}{cc}#1 & #2\vspace{2truemm} \\ #3 & #4
\end{array}\right)}}

\def\tilde{\widetilde}
\def\ri{{\rm{i}}}

\begin{document}

\title[Continuous Symmetries of the lpKdV equation]{\bf Continuous Symmetries of the Lattice Potential KdV Equation}

\author{DECIO LEVI}

\address{Dipartimento di Ingegneria Elettronica \\
Universit\`a degli Studi Roma Tre and Sezione INFN, Roma Tre \\
Via della Vasca Navale 84, 00146 Roma, Italy\\
E-mail: levi@fis.uniroma3.it}

\author{MATTEO PETRERA} 

\address{Zentrum Mathematik \\
Technische Universit\"at M\"unchen \\
Boltzmannstr. 3,
D-85747 Garching bei M\"unchen,
Deutschland\\
Dipartimento di Fisica E. Amaldi \\
Universit\`a degli Studi Roma Tre and Sezione INFN, Roma Tre \\
Via della Vasca Navale 84, 00146 Roma, Italy\\
E-mail: petrera@ma.tum.de}

\begin{abstract}
In this paper we present a set of results on the integration and on the symmetries of the lattice potential 
Korteweg-de Vries (lpKdV) equation. Using its associated spectral problem we construct
the soliton solutions and the Lax technique enables us to provide   infinite sequences of generalized symmetries. 
Finally,  using a discrete symmetry of the lpKdV equation, we construct a
large class of non-autonomous  generalized symmetries.  
\end{abstract}

\maketitle

\section{Introduction}

The lattice version of the potential Korteweg-de Vries (lpKdV) equation
\beq \label{pkdv}
w_t=w_{xxx}+3 \,w_x^2,
\eeq
is given by the nonlinear partial difference equation \cite{frank}:
\beq
\mathbb{D} \doteq (p-q + u_{n,m+1} - u_{n+1,m})\,(p+q - u_{n+1,m+1} + u_{n,m})-(p^2-q^2) = 0.\label{kdv}
\eeq
The above equation is probably the best known completely discrete equation which involves just four 
points which lay on two
orthogonal infinite lattices and are the vertices of an elementary square
- a quad-graph - (see Fig.1). 
Eq. (\ref{kdv}) is one of the lattice equations on quad-graphs classified in \cite{abs}, where the $3D$ {\it consistency} is used as a tool to establish
its integrability.

\vspace{.5cm}
\begin{center}
\begin{figure}[h]
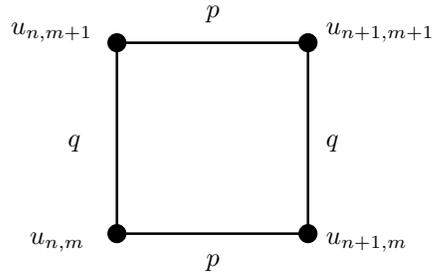

\centertexdraw{\setunitscale 0.5
\linewd 0.03 \arrowheadtype t:F
\move (-1 -1)
\lvec (-1 1) \lvec (1 1) \lvec (1. -1) \lvec(-1 -1)
\move(-1 -1) \fcir f:0.0 r:0.1 \move(-1 1) \fcir f:0.0 r:0.1
\move(1 1) \fcir f:0.0 r:0.1 \move(1 -1) \fcir f:0.0 r:0.1
\htext (-1.9 -1.2) {$u_{n,m}$}
\htext (-2.1 1.) {$u_{n,m+1}$}
\htext (1.2 1) {$u_{n+1,m+1}$}
\htext (1.2 -1.2) {$u_{n+1,m}$}
\htext (-0.05 1.2) {$p$}
\htext (-0.05 -1.4) {$p$} 
\htext (-1.5 -0.15) {$q$}
\htext (1.2 -0.15) {$q$}
}
\caption{An elementary square}
\end{figure}
\end{center}

The lpKdV equation has been introduced for the first time by Hirota
\cite{hirota} in 1977 and it is nothing else but 
the nonlinear superposition formula for the Korteweg-de Vries equation 
under disguise. A review of results about the lpKdV equation can be found in \cite{frank}.

Integrable equations possess an infinite set of symmetries. Few of them are {\it point symmetries}, i.e. 
symmetries whose infinitesimal generators depend just on the independent and dependent variables, while an
infinite denumerable number of them 
are {\it generalized symmetries}. The latter ones depend  on the derivatives of the dependent variable with respect to 
 the continuous independent variables and on a few lattice points if the independent variables are discrete. 
The presence of this infinite Lie algebra of symmetries and of the associated conserved quantities is one of the most important features of the
integrability of a given nonlinear equation and it has been used with profit in the past to provide
integrability tests for several partial differential equations in $\RR^2$ and differential-difference
equations \cite{yamilov1,yamilov2}. 

To be able to introduce an integrability test based on symmetry one needs to understand the structure of the 
infinite dimensional symmetry algebra of integrable equations. 
In the case of completely discrete equations the situation is not as clear as 
for the differential-difference or the partial-differential case.
A result in this direction has been obtained
for the discrete-time Toda lattice \cite{martina}.  
However, the Toda lattice is just an example and more examples are needed to get a sufficiently general 
idea of the possible structures which may appear. 

On the other hand, in \cite{JLP, levi, levipetrera}, the
multiscale expansion technique \cite{tk} has been extended to the
case of equations defined on a lattice. In \cite{levi} one of the authors has performed,
as an example, the multiscale expansion of eq. (\ref{kdv}) deriving  a completely
local discrete nonlinear Schr\"odinger equation (dNLS). The integrability
of this dNLS equation has been questioned by many people. So the construction of the
symmetries of the lpKdV equation and their multiscale expansion should be a concrete tool to
check its integrability.

\vspace{.3truecm}

The present paper is devoted to the study of the lpKdV equation exactly with these aims.
In Section \ref{S1} we review some known results on this lattice equation, while in Section \ref{S2} we present its 
inverse scattering transform, which turns out to be a slight generalization of the results presented 
by Boiti and collaborators on an asymmetric discretization of the continuous Schr\"odinger spectral problem \cite{boiti2}. 
Section \ref{S3} is devoted to the construction of Lie point and generalized  symmetries. Finally,
Section \ref{S4} contains some concluding remarks and some open problems.

\section{The lattice potential KdV equation} \label{S1}

Here we  present some known results on the lpKdV equation (\ref{kdv}) and its integrability.  

In eq. (\ref{kdv}) $u_{n,m}$ is the dynamical field  variable, 
which we assume to be real and asymptotically bounded by a constant,  defined
at the site
$(m,n) \in \mathbb{Z}^2$ while 
$p,q$ are two nonzero real parameters which are related to the 
lattice steps and  will 
go to infinity in the continuous limit so as to obtain from eq. (\ref{kdv}) the continuous
potential KdV equation (\ref{pkdv}).

Since we have two discrete independent variables,  $n$ and $m$,
we  perform the continuous limit in two steps. Each step is   achieved by shrinking the
corresponding lattice step to zero and sending to infinity the number of lattice points.

In the first step we
define $u_{n,m} \doteq v_{k}(\tau)$, where
$k \doteq n+m$ and $\tau \doteq \delta \,  m$, being  $\delta \doteq p-q$ the lattice step in the $m$-direction. 
In the limit 
$m \rightarrow \infty, \, \delta \rightarrow 0$, we obtain the
differential-difference equation
\beq
\frac{\partial v_{k}}{\partial \tau}=\frac{2\,p}{2\,p-v_{k+1}+v_{k-1}}-1. \label{2}
\eeq
Defining 
\beq \nonumber
q_k \doteq 2\,p-v_{k+2}+v_{k},
\eeq
we can rewrite eq. (\ref{2}) as the
 differential-difference equation
\beq
\frac{\partial q_{k}}{\partial \tau}=2\,p \left(\frac{1}{q_{k-1}} - \frac{1}{q_{k+1}}\right). \label{3}
\eeq
Eq. (\ref{3}) has  been obtained in ref. \cite{boiti} as the simplest local evolution equation associated with
the asymmetric discrete Schr\"odinger spectral problem 
\beq \label{sp}
\psi_{k+2}  = q_k \,\psi_{k+1} + \lambda \, \psi_k,
\eeq
where $\l \in \mathbb{C}$ is a spectral parameter.

Introducing the new field $s_k \doteq (2\, p)/q_k$, eq. (\ref{3}) reads
\beq \label{dkdv}
 \frac{\partial s_{k}}{\partial  \tau}= s_k^2 \,(s_{k+1}-s_{k-1}),
\eeq
the so called {\it discrete KdV equation}  \cite{narita}.
Then the Miura transformation   $a_k \doteq s_k \, s_{k-1}$, maps eq.
(\ref{dkdv})  into the {\it discrete Volterra equation}
\beq \label{vol}
 \frac{\partial a_{k}}{\partial  \tau}= a_k \,(a_{k+1}-a_{k-1}),
\eeq
associated with the Schr\"odinger spectral problem
\beq \nonumber
\psi_{k-1}  + a_k \,\psi_{k+1} = \mu \, \psi_k,
\eeq
where $\mu \in \mathbb{C}$ is a spectral parameter.

The second continuous limit of eq. (\ref{kdv}) is performed by defining $v_{k}(\tau)\doteq w(x,t)$, with 
$
x\doteq 2\, \left(k + \tau/p   \right)/p$ and
$t\doteq 2 \, \left(k/3 + \tau /p \right)/p^3.
$
If we carry out the limit 
$
p \rightarrow \infty, \, k \rightarrow \infty, \, \tau \rightarrow \infty,
$ in such a way that $x$ and $t$ remain finite, then  
eq. (\ref{2}) is transformed into the  potential KdV equation  (\ref{pkdv}).

The integrability of the lpKdV equation (\ref{kdv}) is proven in ref. \cite{frank} by giving its Lax pair, 
an overdetermined system of matrix 
equations for the vector $\Psi^h_{n,m} \doteq (\psi^1_{n,m}(h),\psi^2_{n,m}(h))^T$, where $h \in \mathbb{C}$ is a  spectral parameter:
\begin{subequations} \label{lax12}
\begin{align}
&\Psi^h_{n+1,m}=L_{n,m}^h \, \Psi^h_{n,m}, \label{lax1} \\
&\Psi^h_{n,m+1}=M_{n,m}^h \, \Psi^h_{n,m}, \label{lax2}
\end{align}
\end{subequations}
with
$$
L_{n,m}^h \doteq \left(\begin{array}{cc}
p-u_{n+1,m} & 1  \\
h^2-p^2+(p+u_{n,m})\,(p-u_{n+1,m}) & p+u_{n,m}
\end{array}\right),
$$
and
$$
M_{n,m}^h \doteq \left(\begin{array}{cc}
q-u_{n,m+1} & 1  \\
h^2-q^2+(q+u_{n,m})\,(q-u_{n,m+1}) & q+u_{n,m}
\end{array}\right).
$$
The consistency of eqs. (\ref{lax12}) implies the discrete Lax equation 
\beq \nonumber
L_{n,m+1}^h \, M_{n,m}^h=M_{n+1,m}^h \,L_{n,m}^h,
\eeq
which plays the same role as the zero-curvature equation $L_{,t} = [ M \, , L \,] $
in the continuous case.  The lpKdV equation (\ref{kdv}) corresponds to an isospectral deformation of the Lax pair
(\ref{lax12}), i.e. whenever $h$ is an $m$ and $n$-independent complex constant. We can rewrite 
eqs. (\ref{lax12}) in  scalar form in terms of $\psi_{n,m} \doteq \psi^1_{n,m}(h)$:
\begin{subequations}
\begin{align}
&\psi_{n+2,m}=(2\,p-u_{n+2,m}+u_{n,m})\, \psi_{n+1,m}+(h^2-p^2)\, \psi_{n,m}, \label{lax3}\\
&\psi_{n,m+1}=\psi_{n+1,m}+(q-p+u_{n+1,m}-u_{n,m+1}) \,\psi_{n,m}.\label{lax4}
\end{align}
\end{subequations}
Taking into account eqs. (\ref{lax4}--\ref{kdv}) we can
rewrite the $m$-evolution also as
\beq \label{lax4a}
\psi_{n,m+2}=(2\,q-u_{n,m+2}+u_{n,m})\, \psi_{n,m+1}+(h^2-q^2)\, \psi_{n,m}.
\eeq
Eq. (\ref{lax4a}) coincides with eq. (\ref{lax3}) swapping the indices
$n \leftrightarrow m$ and $p \leftrightarrow q$. It is easy to see that also the lpKdV equation  (\ref{kdv})
has the same discrete symmetry.

Defining
\beq \label{qn}
q_{n} \doteq q_{n,m} \doteq 2\,p-u_{n+2,m}+u_{n,m},
\eeq
choosing $\lambda \doteq h^2 - p^2$ and dropping the parametric dependence on $m$ we see that eq. (\ref{lax3}) is equivalent
to the discrete Schr\"odinger spectral problem (\ref{sp}).
The $m$-evolution,  provided by eq. (\ref{lax4}),  cannot be written in a simple way in terms of $q_{n,m}$.

\section{The discrete spectral problem associated with the lpKdV equation} \label{S2}

Let us now study
the direct and inverse problems associated with eq. (\ref{sp}).  Our results are
a  generalization of those contained in \cite{boiti2}: in fact they reduce to them when
$p=1$. Henceforth we remand to  \cite{boiti2} for most of the technical
details. 

As the independent discrete variable $m$ 
- the discrete time variable of the lpKdV equation - enters  parametrically in the results of this section, 
 it will not be explicitly written.
We shall write it just when it is necessary, namely when we will study the $m$-evolution of the 
spectral data associated with the lpKdV equation.

\subsection{Direct problem}

The solutions $u_{n,m}$ of eq. (\ref{kdv}) must go asymptotically to an arbitrary constant 
to be consistent with the difference equation (\ref{kdv}).  
Then $q_n$, defined in eq. (\ref{qn}),  goes asymptotically  to $ 2\, p$.

Following \cite{boiti2} we define
\begin{subequations} \label{tr12}
\begin{align}
& h \doteq {\rm i}\, \kappa, \qquad \qquad \; \; \; \lambda \doteq -\kappa^2-p^2, \label{tr1} \\
&  q_n \doteq \eta_n+2\,p, \qquad \psi_n \doteq (p+ \ri \,\kappa)^n \,\chi_n, \label{tr2} 
\end{align}
\end{subequations}
where $\eta_n \doteq -u_{n+2,m}+u_{n,m}$ is a new asymptotically vanishing potential.
In terms of the new variables  (\ref{tr12}) the spectral problem (\ref{sp}) reads
\beq
(p +\ri \,\kappa) \,\chi_{n+2} -2\,p \,\chi_{n+1}+(p- \ri\, \kappa)\, \chi_n = \eta_n \,\chi_{n+1}.\label{sp2}
\eeq

The Jost solutions  $\mu_n^{\pm}$ of the spectral problem (\ref{sp2})  
are defined  in terms of the potential $\eta_n$ and of the discrete complex exponential  
$[(p+\ri \,\kappa)/(p- \ri\, \kappa)]^n$ through the following ``discrete integral equations''
\begin{subequations} \label{jost12}
\begin{align}
&\mu^+_n = 1 - \frac{1}{2\,\ri \,k}
\sum_{j=n+1}^{+ \infty} \left[ 1 +\left(\frac{p+ \ri\, \kappa}{p - \ri\, \kappa}\right)^{j-n}\right]\,
\eta_{j-1} \,\mu^+_j, \label{jost1}\\
&\mu^-_n = 1 + \frac{1}{2\, \ri \,\kappa}
\sum_{j=-\infty}^{n} \left[ 1 +\left(\frac{p+ \ri \,\kappa}{p - \ri\, \kappa}\right)^{j-n}\right]\,
\eta_{j-1} \,\mu^-_j. \label{jost2}
\end{align}
\end{subequations}
The Jost solution $\mu_n^+$ is 
an analytic function of $\kappa$ for 
${\rm {Im}}(\kappa) > 0$ and $\mu_n^-$ for ${\rm {Im}}(\kappa) < 0$ with the boundary conditions 
\begin{subequations} \label{l12}
\begin{align}
\lim_{n \rightarrow + \infty} \mu^+_n =1 , \qquad {\rm {Im}}(\kappa) \geq 0,  \label{l1}\\
\lim_{n \rightarrow - \infty} \mu^-_n =1 , \qquad {\rm {Im}}(\kappa) \leq 0. \label{l2}
\end{align}
\end{subequations}

For ${\rm {Im}}(\kappa)=0$ we define the spectral data:
\begin{subequations}
\begin{align}
&a^\pm(\kappa) \doteq 1 \mp \frac{1}{2\, \ri \,\kappa}
\sum_{j=-\infty}^{+ \infty} \eta_{j-1} \,\mu^\pm_j,  \nonumber \\
&b^\pm(\kappa)\doteq  \pm \frac{1}{2\, \ri\, \kappa}
\sum_{j=-\infty}^{+ \infty} \left(\frac{p+ \ri\, \kappa}{p - \ri \,\kappa}\right)^{j}\eta_{j-1} \,\mu^\pm_j.  \nonumber
\end{align}
\end{subequations}
Due to the analiticity property of the
Jost solutions one can prove that  $a^+(k)$ can be analitically extended 
to ${\rm {Im}}(\kappa) > 0$ and $a^-(\kappa)$ to ${\rm {Im}}(\kappa) < 0$. The inverse of the
functions $a^\pm(\kappa)$, namely $T^\pm(\kappa) \doteq [a^\pm(\kappa)]^{-1}$, play the role of the 
{\it transmission coefficients} 
and the functions
$
R^\pm(\kappa) \doteq b^\pm(\kappa)/ a^\pm(\kappa)
$
are the {\it reflection coefficients}. The poles of $T^\pm(\kappa)$  are related
to the soliton solutions of the evolution equations associated with the spectral problem (\ref{sp2}). 
Assuming that eqs. (\ref{jost12}) are
solvable and their solution is  unique, we obtain for ${\rm {Im}}(\kappa) = 0$ the following relations between the 
Jost solutions and the spectral data,
\beq
\mu^\pm_n(\kappa)= a^\pm(\kappa) \,\mu^\mp_n(\kappa) + \left(\frac{p- \ri\, \kappa}{p + \ri \,\kappa}\right)^n
 b^\pm(\kappa) \,\mu^\mp_n(-\kappa). \nonumber
\eeq
Taking into account the asymptotic limits (\ref{l12}) of the Jost functions, for $ {\rm {Im}}(\kappa) = 0$ we get
\beq
\mu^\pm_n \sim [T^\pm(\kappa)]^{-1} \left[1 + 
\left(\frac{p- \ri\, \kappa}{p + \ri\, \kappa}\right)^n R^\pm(\kappa) \right], \quad
n \rightarrow \mp \infty \label{mu}.
\eeq
Let us assume that the function $a^+(\kappa)$ has $N$ simple zeros at
$\kappa=\kappa_j^+,\, 1 \leq j \leq N$. In correspondence with these values of
$\kappa$ the Jost function is asymptotically bounded and 
the $N$ residues of the transmission coefficient $T^+(\kappa)$, say
$\{c_j^+\}_{j=1}^N$, are associated with its normalization coefficient.

Henceforth  we can associate  with a given potential $\eta_n$ in a unique way the spectral data 
$\Ss [ \eta_{n} ]\doteq $ $\{R(\kappa), T(\kappa), \{c_j\}_{j=1}^N\}$, obtained 
as a solution of the spectral problem (\ref{sp2}). 
As the spectral problem (\ref{sp2}) is a linear ordinary difference equation in $n$, 
the solution $\chi_{n}$ is defined up to an arbitrary normalization function $\Omega_m(\kappa)$.

\subsection{Inverse problem}

Taking
into account eqs. (\ref{l12}--\ref{mu}) and using the Cauchy-Green formula we can write the following integral equations for the two Jost solutions
\begin{subequations} \label{cg12}
\begin{align}
T(\kappa) \,\mu^+_n(\kappa)=&\, 1 + \sum_{j=1}^N C_j \frac{\mu^-_n(-\kappa_j^+)}{\kappa-\kappa_j^+}
\left(\frac{p- \ri \,\kappa_j^+}{p + \ri\, \kappa_j^+}\right)^n+ \nonumber \\
& + \frac{1}{2 \, \ri \,\kappa} \int_{-\infty}^{+\infty}\frac{\mu^-_n(-s) \,R(s)}{s-\kappa - \ri\, 0}
\left(\frac{p- \ri\, s}{p + \ri\, s}\right)^n ds, \qquad {\rm {Im}}(\kappa) \leq 0, \label{cg1} \\
& {} \nonumber \\
\mu^-_n(\kappa)=&\, 1 + \sum_{j=1}^N C_j \frac{\mu^-_n(-\kappa_j^+)}{\kappa-\kappa_j^+}
\left(\frac{p- \ri\, \kappa_j^+}{p + \ri\, \kappa_j^+}\right)^n+ \nonumber \\
& + \frac{1}{2\,  \ri \,\kappa} \int_{-\infty}^{+\infty}\frac{\mu^-_n(-s)\,R(s)}{s-\kappa + \ri \,0}
\left(\frac{p- \ri\, s}{p + \ri \,s}\right)^n ds, \qquad {\rm {Im}}(\kappa) \geq 0, \label{cg2}
\end{align}
\end{subequations}
where
\beq \label{cg123}
C_j \doteq \frac{c_j^+}{2\, \ri\, \kappa_j^+}\sum_{k=-\infty}^{+\infty}
\left(\frac{p+ \ri\, \kappa_j^+}{p - \ri \,\kappa_j^+}\right)^k \eta_{k-1}\, \mu^+_k(\kappa_j^+).
\eeq
Therefore, given the spectral data $\Ss [\eta_n]$,  we are able to
reconstruct the Jost solutions and, in terms of them, the potential
\bea \nonumber 
&& \back \back \eta_{n,m} = \ri \sum_{j=1}^N C_{j} \,\left[ \mu^-_{n+2,m}(-\kappa_j^+) 
\left(\frac{p+ \ri\, \kappa_j^+}{p - \ri \,\kappa_j^+}\right)^2 -
\mu^-_{n,m}(-\kappa_j^+)\right]
 \left(\frac{p+ \ri\, \kappa_j^+}{p - \ri \,\kappa_j^+}\right)^n- \\  
&&   \qquad - \int_{-\infty}^{+\infty} \left[ \mu^-_{n+2,m}(-s) \left(\frac{p+ \ri\, s}{p
        - \ri \,s}\right)^2 - \mu^-_{n,m}(-s) \right]
  \left(\frac{p+ \ri\, s}{p - \ri \,s}\right)^n R_m^+(s) \, ds.\label{cg1234}
\eea 
We remand to \cite{boiti2} for further details and for a study of the convergence of
the sum appearing in the definition (\ref{cg123}) of the coefficients
$C_j$.

\subsection{Time evolution of spectral data}

Taking
into account the definitions  (\ref{tr12}), 
the $m$-evolution of  $\chi_{n,m}$ is obtained from eq. (\ref{lax4}) and is given by
\beq 
\chi_{n,m+1}=(p + \ri \,\kappa) \,\chi_{n+1,m}+(q-p+u_{n+1,m}-u_{n,m+1})\,\chi_{n,m}. \label{asy22}
\eeq
As the lpKdV equation (\ref{kdv}) is an isospectral deformation of eq. (\ref{sp2}), from (\ref{tr1})  $\kappa$ is an $m$-independent parameter.

Defining
$\chi_{n,m} \doteq \Omega_m(\kappa) \, \mu^+_{n,m}$,
where $\Omega_m(\kappa)$ is a normalization function,
and taking into account the asymptotic behaviours (\ref{l12}--\ref{mu})  of the Jost functions, we find from eq. (\ref{asy22}) the following  
time evolution of $\Ss[\eta_{n,m}]$
\beq \label{eqab}
T_{m+1}(\kappa) = T_m(\kappa),\qquad R_{m+1}(\kappa) = \left(\frac{q- \ri \,\kappa}{q+\ri \,\kappa}\right)  \, R_m(\kappa).
\eeq
Integrating eqs. (\ref{eqab}) we get
\beq \label{eqabc}
T_m(\kappa)= T(\kappa), \qquad
R_m(\kappa)= \left(\frac{q -\ri \, \kappa}{q +\ri\, \kappa} \right)^m  R(\kappa), \nonumber
\eeq
namely the transmission coefficient is invariant under the evolution of the lpKdV equation (\ref{kdv}),
while the reflection coefficient adquires an $m$-dependent discrete exponential factor.

\subsection{Soliton solutions}

The soliton solutions of eq. (\ref{kdv}) are obtained by considering  the
reflectionless solution of the spectral problem (\ref{lax3}--\ref{lax4}--\ref{tr12}) . 
They are obtained from eq. (\ref{cg1234}) setting $R^+_m(\kappa)=0$. The Jost
solutions $\mu_{n,m}^-(-\ri \, \kappa^+_j)$ are obtained solving eq.
(\ref{cg2}) for $\kappa = -\kappa^+_j$. For $N=1$ they are given by
\beq \nonumber
\mu_{n,m}^-(-\kappa^+) = \left [1+\frac{C_m}{2 \, \kappa^+} \left(\frac{p-
      \ri\, \kappa^+}{p + \ri\, \kappa^+}\right)^n \right]^{-1},
\eeq
while for $N=2$ they read
\bea \nonumber
&& \mu_{n,m}^-(-\kappa_1^+) = \frac{1}{\mathcal D_{1,2}}
\left[ 1+\frac{C_{2,m}}{ \kappa_2^+-\kappa_1^+}
  \left(\frac{p- \ri\, \kappa_2^+}
{p + \ri\, \kappa_2^+}\right)^n \right], \\ \nonumber
&& \mu_{n,m}^-(-\kappa_2^+) = \frac{1}{\mathcal D_{1,2}}
\left[ 1+\frac{C_{1,m}}{ \kappa_1^+-\kappa_2^+}
  \left(\frac{p- \ri\, \kappa_1^+}
{p + \ri\, \kappa_1^+}\right)^n \right],
\eea
where
\bea \nonumber
\mathcal D_{1,2} &\doteq&  1+\frac{C_{2,m}}{2 \, \kappa_2^+} \left(\frac{p- \ri\,
    \kappa_2^+}{p + \ri\, \kappa_2^+}\right)^n 
+ \frac{C_{1,m}}{2 \, \kappa_1^+} \left(\frac{p- \ri\, \kappa_1^+}{p + \ri\,
    \kappa_1^+}\right)^n + \\ \nonumber 
&&  + \, \frac{C_{1,m} \, C_{2,m} ( \kappa_2^+-\kappa_1^+)^2}{ 4 \,
  \kappa_1^+ \, 
  \kappa_2^+(\kappa_2^++\kappa_1^+)^2}  
\left(\frac{p- \ri\, \kappa_1^+}{p + \ri\, \kappa_1^+}\right)^n \left(\frac{p- \ri\, \kappa_2^+}{p + \ri\, \kappa_2^+}\right)^n.
\eea
Taking into account the relation between $\eta_{n,m}$ and $u_{n,m}$ and the
time evolution (\ref{eqab}) 
we get that the one soliton solution for the lpKdV equation (\ref{kdv}) is given by
\beq 
 u_{n,m}= -\frac{\ri\,  C \left(\frac{p-
  \ri\, \kappa^+}
{p + \ri\, \kappa^+}\right)^n \left(\frac{q- \ri\, \kappa^+}{q + \ri\,
  \kappa^+}\right)^m }{
1+\frac{C}{2 \, \kappa^+} \left(\frac{p- \ri\,
       \kappa^+}{p + \ri\, \kappa^+}\right)^n 
\left(\frac{q- \ri\, \kappa^+}{q + \ri\, \kappa^+}\right)^m }. \label{solkdv}
 \eeq
 By proper choices of the complex constants $C$ and $\kappa^+$, where
$\kappa^+$ is defined in the upper half plane, and of the real constants $p$ and $q$ one can
always render the solution (\ref{solkdv}) real. 

The two soliton solution is given by
\beq 
 u_{n,m}= -\frac{\ri}{\mathcal D_{1,2}} \sum_{j=1}^2 \frac{  C_j \left(\frac{p-
  \ri\, \kappa_j^+}
{p + \ri\, \kappa_j^+}\right)^n \left(\frac{q- \ri\, \kappa_j^+}{q + \ri\,
  \kappa_j^+}\right)^m }{
1+\frac{C_j}{2 \, \kappa_j^+} \left(\frac{p- \ri\,
       \kappa_j^+}{p + \ri\, \kappa_j^+}\right)^n 
\left(\frac{q- \ri\, \kappa_j^+}{q + \ri\, \kappa_j^+}\right)^m }. \nonumber
 \eeq
with $\mathcal D_{1,2}$ given above.

\section{Continuous symmetries of the lpKdV equation} \label{S3}

Lie symmetries of the lpKdV equation (\ref{kdv}) are given by those continuous
transformations which leave the equation invariant. From the infinitesimal point of view 
they are obtained by requiring the infinitesimal invariant condition 
\beq \label{cca2}
\left. {\rm pr} \, \widehat X_{n,m} \, \mathbb{D}  \, \right|_{\mathbb{D} =0} =0,
\eeq
where 
\beq \label{ccb2}
 \widehat X_{n,m} = F_{n,m} ( u_{n,m}, u_{n \pm 1,m}, u_{n, m \pm 1}, \ldots) \partial_{u_{n,m}}.
 \eeq
By $ {\rm pr} \, \widehat X_{n,m}$ we mean the prolongation of the infinitesimal generator 
$\widehat X_{n,m}$  to all points appearing in $\mathbb{D}=0$.

If $F_{n,m} = F_{n,m}( u_{n,m})$ then we get {\it point symmetries} and the procedure to get them from eq. (\ref{cca2})
is purely algorithmic \cite{lw6}. 
{\it Generalized symmetry}  are obtained when $F_{n,m} = F_{n,m} ( u_{n,m}, u_{n \pm 1,m}, u_{n, m \pm 1}, \ldots)$. 
In the case of nonlinear discrete equations,
the Lie point symmetries are not very common, but, if the equation is integrable and there exists 
a Lax pair, it is possible to construct an infinite family of generalized symmetries. 

In correspondence with the infinitesimal generator (\ref{ccb2}) 
we can in principle construct 
a group transformation by integrating the initial boundary problem
\beq \label{s1}
\frac{d \tilde u_{n,m}(\ep)}{d \ep} = F_{n,m} ( \tilde u_{n,m}(\ep), 
\tilde u_{n \pm 1,m}(\ep), \tilde u_{n, m \pm 1}(\ep), \ldots),
\eeq
with $\tilde u_{n,m}(\ep =0) = u_{n,m}$, where   $\ep \in \mathbb{R}$ is the
continuous Lie group parameter.
This can be done effectively only in the case of point symmetries as in the
generalized 
case we have a differential-difference equation for which we cannot find the  solution for a generic initial data, but, at most, we can find some particular solutions. 
 Eq. (\ref{cca2}) is equivalent to the request that the $\ep$-derivative of the equation $\mathbb{D}=0$, 
written for $\tilde u_{n,m}(\ep)$, 
is identically satisfied when the $\ep$-evolution of $\tilde u_{n,m}(\ep)$ is given by eq. (\ref{s1}).
This is also equivalent to say  that the flows (in the group parameter space) given by eq. (\ref{s1}) 
are compatible or commute with $\mathbb{D}=0$.

By means of a symmetry transformation we can construct an $\ep$-dependent class of 
solutions of eq. (\ref{kdv}), 
given by the functions $\tilde u_{n,m} (\ep)$. We can associate with this
class
of solutions the corresponding solution $\tilde \psi_n$ of the spectral
problem (\ref{lax3}) and eq. 
(\ref{s1}) can be found among those nonlinear evolution equations associated with the spectral problem 
(\ref{lax3}) which are commuting with eq. (\ref{kdv}).

\subsection{Lie point symmetries}

As we do not want to change the basic lattice structure of our space, we just
consider transformations $G_\ep$ acting on the domain of the dependent variables:
$$
G_\ep: \, u_{n,m} \,\mapsto \, \Phi_{n,m}  (u_{n,m}; \ep), \qquad \ep \in \mathbb{R}.
$$
The infinitesimal generator of the group action of $G_\ep$
on $u_{n,m}$ is the vector field
\beq \nonumber
\widehat X_{n,m} = \phi_{n,m}(u_{n,m}) \, \partial_{u_{n,m}},
\eeq
where
$$
\phi_{n,m}(u_{n,m})  \doteq \left.
\frac{ d}{{d} \varepsilon}
\Phi_{n,m}  (u_{n,m}; \ep) \right|_{\varepsilon=0} .
$$
There is a one-to-one correspondence between connected groups of
transformations and their associated infinitesimal generators  \cite{olver}. The group
action of $G_\ep$ can be reconstructed from the flow of the infinitesimal vector field $\widehat X_{n,m}$ either by
exponentiation:
$$
\Phi_{n,m}  (u_{n,m}; \ep)=\exp (\varepsilon \,\widehat X_{n,m} ) \,u_{n,m},
$$
or by solving the  differential boundary value problem
\beq \label{cc1}
\frac{d \tilde u_{n,m}(\ep)}{d \ep} = \phi_{n,m}(\tilde u_{n,m}(\ep)), 
\eeq
with the initial condition $ \tilde u_{n,m}(\ep=0)=u_{n,m}$.
Eq. (\ref{cc1}) is just a subcase of eq. (\ref{s1}) when $F_{n,m}$ depends just on $u_{n,m}$.
The prolongation of the infinitesimal action of $G_\ep$ is given by 
\bea
\qquad {\rm pr} \, \widehat X_{n,m} & =& \phi_{n,m}(u_{n,m})\,\partial_{u_{n,m}} +
\phi_{n+1,m}(u_{n+1,m})\,\partial_{u_{n+1,m}} + \label{prps1} \\
&& + \,  \phi_{n,m+1}(u_{n,m+1})\,\partial_{u_{n,m+1}} +
\phi_{n+1,m+1}(u_{n+1,m+1})\,\partial_{u_{n+1,m+1}}. \nonumber
\eea

Applying the prolongation (\ref{prps1}) to eq. (\ref{kdv}) we get
\bea \label{ps2}
&& \left. \qquad {\rm pr} \, \widehat X_{n,m} \, \mathbb{D} \right|_{\mathbb{D} =0} = \\
&&\back \back=[\phi_{n,m}(u_{n,m}) - \phi_{n+1,m+1}(u_{n+1,m+1})]\, (p - q + u_{n,m+1} - u_{n+1,m} ) + \nonumber \\
&&\back \back +\, 
( p + q - u_{n+1,m+1} + u_{n,m} ) \,[ \phi_{n,m+1}(u_{n,m+1}) - \phi_{n+1,m}(u_{n+1,m})] 
\bigg|_{\mathbb{D} =0} = 0. \nonumber
\eea
Due to eq. (\ref{kdv})  just three of the 
four different fields appearing in eq. (\ref{ps2}) are independent and, without loss of generality, we choose them to be 
$u_{n,m}$, $u_{n,m+1}$ and $u_{n+1,m}$.

From eq. (\ref{ps2}), by differentiation with respect to $u_{n,m}$, we  obtain, when $p \neq q$,
\beq \label{ps2a}
\frac{d\phi_{n,m}}{d u_{n,m}} = \frac{d \phi_{n+1,m+1}}{d u_{n+1,m+1}},
\eeq
so that
\beq
\frac{d^2 \phi_{n,m}}{d u_{n,m}^2} = 0 \qquad \Rightarrow \qquad
\phi_{n,m} = \mathcal K^0_{n,m} +  \mathcal K^1_{n,m} u_{n,m},\label{ps3}
\eeq
where $\mathcal K^0_{n,m}$ and $\mathcal K^1_{n,m}$ are two integration constants depending just upon $n$ and $m$.
Inserting  $\phi_{n,m}$ given in eq. (\ref{ps3}) into eq. (\ref{ps2a}) we find that 
$\mathcal K^1_{n,m} =  \mathcal K^1_{n+1,m+1}$, while if we insert it into
eq. (\ref{ps2}) we get an explicit equation  in $u_{n,m}$, $u_{n,m+1}$ and $u_{n+1,m}$.
Then the various powers of $u_{n,m}$, $u_{n,m+1}$ and $u_{n+1,m}$ give us a set of coupled linear partial difference 
equations for $ \mathcal K^0_{n,m}$ and $ \mathcal K^1_{n,m}$, whose solutions are:
\begin{subequations}
\begin{align}
& \mathcal K^0_{n,m} = \gamma_1 + (-1)^{n+m} [\,\gamma_2- (q \, m + p \, n) \, \gamma_3\,], \nonumber \\
& \mathcal K^1_{n,m} = (-1)^{n+m}\gamma_3, \nonumber
\end{align}
\end{subequations}
where $\gamma_1,\gamma_2,\gamma_3$ are arbitrary constants.
Therefore the lpKdV equation (\ref{kdv}) admits a three dimensional group $G_\ep$ of Lie point symmetries,
 whose infinitesimal generators are 
\begin{subequations}\label{ps5}
\begin{align}
& \widehat X^1_{n,m} = \partial_{u_{n,m}}, \label{ps5a} \\
& \widehat X^2_{n,m} = (-1)^{n+m} \,\partial_{u_{n,m}}, \label{ps5b} \\
& \widehat X^3_{n,m} = (-1)^{n+m} [ \,u_{n,m} - (p\,n+q\, m)\, ] \, \partial_{u_{n,m}}. \label{ps5c} 
\end{align}
\end{subequations}
The commutation relations between the generators $\widehat X^1_{n,m} ,\widehat X^2_{n,m} ,\widehat X^3_{n,m} $ are:
$$
\left[\,\widehat X^1_{n,m} ,\widehat X^2_{n,m}\, \right]= 0, \quad
\left[\,\widehat X^1_{n,m} ,\widehat X^3_{n,m}\, \right]= \widehat X^2_{n,m}, \quad
\left[\,\widehat X^2_{n,m} ,\widehat X^3_{n,m}\, \right]= \widehat X^1_{n,m}, 
$$
so that they span a solvable Lie algebra.

The finite transformation generated by the symmetry generators (\ref{ps5}) is given by:
\bea
&& \tilde u_{n,m} = e^{\ep_3 \, (-1)^{n+m}} u_{n,m}+ \label{rrr} \\
 && \qquad \quad \; + \, \frac{e^{\ep_3 \, (-1)^{n+m}} -1 }{\ep_3 \, (-1)^{n+m}} \, [\, \ep_1+\ep_2 \, (-1)^{n+m} -\ep_3 \, (-1)^{n+m}(p\, n +q\, m) \, ], \nonumber
\eea
where $\ep_1,\ep_2,\ep_3$ are the group parameters associated respectively
with the symmetry generators $\widehat X^1_{n,m} ,\widehat X^2_{n,m} ,\widehat X^3_{n,m}$.

When $p=q$ eq. (\ref{kdv})
simplifies
to the product of two linear discrete wave equations:
$$
(u_{n,m}- u_{n+1,m+1}+ 2 \, p)\,(u_{n+1,m}- u_{n,m+1})=0,
$$
and the symmetry group becomes infinite dimensional.

Finally, let us recall that eq. (\ref{kdv}) admits  also
some  {\it discrete symmetries}:
\begin{enumerate}
\item The exchange of $n$ and $m$ 
together with the exchange of $p$ and $q$ leaves the lpKdV equation invariant.
\item The exchange of $n$ and $n+1$ 
together with the transformation of $p$ into $-p$ leaves the lpKdV equation invariant.
\item The exchange of $m$ and $m+1$ 
together with the transformation of $q$ into $-q$ leaves the lpKdV equation invariant.
\end{enumerate}

\subsection{Generalized symmetries}

A generalized symmetry is obtained when the function $F_{n,m}$ appearing in eq. (\ref{s1}) 
depends effectively on $u_{n,m}$ evaluated in some lattice points. A way to obtain it is 
to look at those  differential-difference equations (\ref{s1})
associated with the spectral problem (\ref{lax3}) which are compatible with eq.  (\ref{kdv}).
If the  nonlinear differential-difference equation (\ref{s1}) is associated with  
a spectral problem,
 the proof that it is a symmetry of eq. (\ref{s1})  
can be greatly simplified. In fact in this case the spectral problem  establishes  a one-to-one correspondence between eq. (\ref{s1}) and the spectral data and thus one can
  prove 
the commutativity of the flows  in the space of the spectral data, 
where the equations are linear. 

With this aim,  in the following, we construct the hierarchies of nonlinear equations
associated with the spectral problem (\ref{lax3}) of the lpKdV equation (\ref{kdv}). To do so, 
let us rewrite eq. (\ref{lax3}) as an operator equation
\beq \label{s2}
L_n \, \psi_n = \lambda \, \psi_n, \qquad L_n \doteq E_n^{2} - (\eta_{n,m} + 2\, p) \,E_n,
\eeq
where $E_n^{\pm s} \,\psi_n \doteq \psi_{n \pm s}$, $s \in \mathbb{N}$, are the finite shift operators in the $n$ variable
the eigenvalue $\lambda \in \mathbb{C}$ is defined in eq. (\ref{tr1}) 
and $\eta_{n,m} \doteq u_{n,m} - u_{n+2,m}$ is an asymptotically  vanishing potential. 
Starting from eq. (\ref{s2}) we can apply the {\it Lax technique}, 
to obtain the recursion operator $\mathcal L_n$ and the hierarchy of nonlinear evolution equations associated with it, as it has been done in ref.
\cite{boiti}. By the Lax technique we mean that constructive procedure introduced by Bruschi and Ragnisco \cite{br} 
in 1980  which consists in finding  all those operators $M_n$ which are compatible with eq. (\ref{s2}) such that
$
d \psi_n / d \ep = - M_n\,  \psi_n.
$ 
According to the dependence of $\lambda$ on the group parameter $\ep$ we can distinguish between 
two different operator equations for the compatible $L_n$ and $M_n$:
\begin{enumerate}
\item if $d \lambda / d \ep = 0$, then
 \beq \label{s3}
\frac{d L_n}{d \ep} = [ \, \,L_n, M_n \,], \qquad {\rm (isospectral\; hierarchy)};
\eeq
\item if $ d \lambda / d \ep \neq 0$, i.e. $ d \lambda / d \ep = f(\lambda,\epsilon)$, then
\beq \label{s3a}
\frac{d  L_n}{d \ep} = [ \, L_n, M_n\, ] + f(L_n,\epsilon),  \qquad {\textstyle{\rm (non-isospectral\; hierarchy)}}.
\eeq
\end{enumerate}
Each one of the two cases will provide symmetries for the lpKdV equation (\ref{kdv}).
\subsubsection{The isospectral hierarchy}

The isospectral hierarchy of equations is obtained looking at those differential-difference equations 
whose Lax equation is given by eq. (\ref{s3}). 
As we are considering differential-difference equations,  $M_n$ is a suitable  operator expressed in 
terms of shift operators $\{E_n^{\pm s}\}$.
Taking into account eq. (\ref{s2}), the left hand side of eq. (\ref{s3}) is given by 
$-(d \eta_{n,m}(\ep) / d \ep)\, E_n$, 
i.e. it is an operator of the form $V_n(\ep)\, E_n$, where $V_n(\ep)$ is a scalar function. 
The Lax technique described in \cite{br} allows us to write down  the following relation:
\beq \label{lt1}
\tilde V_n = \mathcal L_n \, V_n + V_n^{(0)},
\eeq
where $\tilde V_n= -(d {  \eta}_{n,m}(\tilde \ep) / d \tilde \ep)$, $V_n^{(0)}$ 
is a given function of $\eta_{n,m}$ and of a certain number of arbitrary constants. In this case $V_n^{(0)}$  reads \cite{boiti}
\bea \nonumber
V_n^{(0)} &\doteq & q_{n,m} \left[ (-1)^n \, a_1+ a_2 \,\left( q_{n,m} + 
2 \,p +2 \sum_{k=1}^\infty (-1)^k \eta_{k,m} \right) \right. +\\ \nonumber
&&\left.  + \, a_3 \,(-1)^n \,\left( q_{n,m} + 4 \,p \,n + 2 \sum_{k=1}^\infty \eta_{k,m}\right) \right]   ,
\eea
with $a_1,a_2,a_3$ constants and with $q_{n,m}$ defined in eq. (\ref{qn}). The recursion  operator $\mathcal L_n$ is defined by \cite{boiti}
\beq \label{s4}
{\mathcal L}_n \doteq - q_{n,m} \, \Delta_n^{(-)} \, (\Delta_n^{(+)})^{-1} \,
q_{n,m}  \, E_n (\Delta_n^{(-)})^{-1} \, (\Delta_n^{(+)})^{-1},
\eeq
where
$
\Delta_n^{(\pm)} \doteq E_n \pm 1.
$
The inverses of $\Delta_n^{(\pm)}$ can be easily computed in the space of functions bounded at infinity and are given by
$$ (\Delta_n^{(-)})^{-1} = 
- \sum_{k=0}^{\infty} E_n^{k}, \qquad  (\Delta_n^{(+)})^{-1} = \sum_{k=0}^{\infty} (-1)^k E_n^{k}.
$$
Choosing $V_n=0$, from eq. (\ref{lt1}) a first equation is given by $\tilde V_n = V_n^{(0)}$.
 Iterating this procedure  we obtain 
the following set of equations
\beq \label{s5}
 \frac{d \eta_{n,m}}{d \ep_k} = {\mathcal L}_n^k  \, V_n^{(0)}, \qquad k \in \mathbb{N}.
\eeq
Eqs. (\ref{s5}) involve always at least a summation  and thus 
are not local. Following \cite{boiti} we can obtain a set of 
local equations if we consider in eq. (\ref{s5}) powers of the inverse of the recurrence operator $\mathcal L_n$.
The explicit form of the inverse operator $\mathcal L_n^{-1}$ can be obtained from 
eq. (\ref{s4}) and reads
$$
\mathcal L_n^{-1} = - E_n^{-1} \,\Delta_n^{(-)} \,\Delta_n^{(+)} \,\frac{1}{q_{n,m}} \, 
(\Delta_n^{(-)})^{-1}\, \Delta_n^{(+)}   
\frac{1}{q_{n,m}}.
$$  
Since  $\Delta_n^{(-)} \alpha =0$, where  $\alpha$ is an arbitrary complex constant, 
we get that $(\Delta_n^{(-)})^{-1} 0 = \alpha$ (we can set $\al=1$ without loss of generality). 
So applying $\mathcal L_n^{-1}$ to $0$ we get a local formula for 
$d \eta_{n,m} / d \ep_k$. By a recursive application we find
\beq \label{li1}
\frac{d \eta_{n,m}}{d \ep_k} = \mathcal L_n^{-k-1} \, 0.
\eeq
Eqs. (\ref{li1}) turn out to be local and they  form the so called {\it inverse hierarchy}. 
Taking into account the form of ${\mathcal L}_n^{-1}$ we see that with any equation given in formula (\ref{li1}) 
we can associate an equation for 
the field $u_{n,m}$ given by 
\beq \label{li1a}
\frac{d u_{n,m}}{d \ep_k} = {\tilde {\mathcal L}_n}^{-k} \, \frac{1}{q_{n-1,m}} + \beta_k,
\eeq
where 
\beq \nonumber
\tilde {\mathcal L}_n^{-1} \doteq  - E_n^{-1} \frac{1}{q_{n,m}} 
 \Delta_n^{(+)} (\Delta_n^{(-)})^{-1} \frac{1}{q_{n,m}}  \Delta_n^{(+)} \Delta_n^{(-)}.
\eeq
 Here the $\beta_k$'s  are some integration 
constants to be defined in such a way that $u_{n,m}$, 
asymptotically bounded, is a compatible solution of eq. (\ref{li1a}). 
In correspondence with eqs. (\ref{li1a}) we obtain the following evolution of the reflection coefficient
\beq \label{li12}
\frac{d R_m}{d \ep_k} = \left[ 1 - \left( \frac{p + \ri\, \kappa }{p - \ri\, \kappa} \right)^k \right]  R_m.
\eeq
Taking into account eqs. (\ref{eqab}) and (\ref{li12})  it follows that
\beq \label{li12a}
E_m \, \frac{d R_m}{d \ep_k}  = \frac{d R_{m+1}}{d \ep_k} \qquad \forall \, k \in \mathbb{N},
\eeq
where the $m$-shift operator is defined by $E_m^{\pm s} f_{n,m} \doteq
f_{n,m \pm s}$, $s \in \mathbb{N}$, i.e. eq. (\ref{li1a}) is a symmetry of the
lpKdV equation (\ref{kdv}). 
Moreover, we also have
\beq \label{li12ab}
\frac{d}{d \ep_h} \left( \frac{d R_m}{d \ep_k} \right)= 
\frac{d}{d \ep_k} \left(\frac{d R_m}{d \ep_h}\right) \qquad \forall \, k,h \in \mathbb{N},
\eeq
i.e. the symmetries commute among themselves.
As eq. (\ref{ps5a}) is a symmetry of the lpKdV equation (\ref{kdv}) and the
symmetries span a Lie algebra, also eq. (\ref{li1a}) with $\beta_k =0$ is a  symmetry.
However, when  $\beta_k =0$ the spectral transform is not well defined.

In correspondence with any of the eqs. (\ref{li1a}) with $\beta_k=0$ we can construct an infinitesimal symmetry generator
\beq \label{li1b}
\widehat X_n^{(k)} \doteq \tilde {\mathcal L}^{-k}_n \frac{1}{q_{n-1,m}}\, \partial_{u_{n,m}}.
\eeq
Thanks to eqs. (\ref{li12a}--\ref{li12ab})   the infinitesimal generators (\ref{li1b})
satisfy the symmetry condition (\ref{cca2}) and 
$[\, \widehat X_n^{(k)}, \widehat X_n^{(h)} \, ] = 0$ $\forall \, k,h \in \mathbb{N}$.

For clarity of exposition let us write down the first three equations
of the inverse hierarchy (for simplicity in terms of $q_{n,m}$ given by eq. (\ref{qn})).

For $k=0$ we get
\beq
\frac{d u_{n,m}}{d \ep_0} = -\frac{1}{q_{n-1,m}}+\frac{1}{2 \,p}. \nonumber 
\eeq
For $k=1$ we have
\bea
&& \frac{d u_{n,m}}{d \ep_1}  =  
\frac{1}{q_{n-1,m}^2} \left( \frac{1}{ q_{n,m} } \, + 
\frac{1}{q_{n-2,m} }  \right) - \frac{1}{4\, p^3}. \nonumber 
\eea
For $k=2$ we get
\bea
 \frac{d u_{n,m}}{d \ep_2} &=&  -\frac{1}{q_{n-1,m}^2} \times  \nonumber \\
 && \back  \back \times \, \left[\, \frac{1}{q_{n-2,m}} \,  \left( \frac{1}{q_{n,m} \, q_{n-1,m} }\, + 
 \frac{1}{q_{n-1,m} \, q_{n-2,m} }\, + 
 \frac{1}{q_{n-2,m} \, q_{n-3,m} } \right) \right. +  \nonumber \\
 &&  \back  \back \quad +\left. \frac{1}{q_{n,m}} \,  \left( \frac{1}{q_{n+1,m} \, q_{n,m} }\, + 
 \frac{1}{q_{n,m} \, q_{n-1,m} }\, + 
 \frac{1}{q_{n-1,m} \, q_{n-2,m} } \right) \, \right] + \frac{3}{16\, p^5}. \nonumber
\eea

Before going over to the non-isospectral symmetries let us notice that:
\begin{enumerate}

\item As any of the previous equations can be rewritten just in terms of the variable $q_n$, the mappings $s_n \doteq (2\, p)/q_n$ and $a_n \doteq s_n \, s_{n-1}$
enable us to get the corresponding equations for the discrete KdV and Volterra
equations \cite{hlrw,narita}. 
\item Due to the discrete symmetry $n \leftrightarrow m$, $p \leftrightarrow q$,
there exists a symmetric isospectral hierarchy associated with the spectral problem (\ref{lax4a}) which give rise to
a new denumerable class of commuting symmetries. These symmetries are  given by eq. (\ref{li1a}) with $q_{n,m}$ 
substituted by 
\beq \nonumber
p_{n,m} \doteq 2 \, q - u_{n,m+2} + u_{n,m},
\eeq 
and ($E_n^\pm$, $\Delta_n^{(\pm)}$, $\tilde {\mathcal L}_n$) 
substituted 
by ($E_m^\pm$, $\Delta_m^{(\pm)}$, $\tilde {\mathcal L}_m$). The
infinitesimal symmetry generators of this hierarchy are given by
\beq 
\nonumber
\widehat X_m^{(k)} \doteq \tilde {\mathcal L}_m^{-k} \frac{1}{p_{n,m-1}} \, \partial_{u_{n,m}},
\eeq
which are also commuting among themselves, i.e. 
$[\, \widehat X_m^{(k)}, \widehat X_m^{(h)} \, ] = 0$ $\forall \, k,h \in
\mathbb{N}$.
Denoting by $\tau_k$ the continuous  
group parameter of this sequence, the first two symmetries of this class are thus given by:
\bea
\frac{d u_{n,m}}{d \tau_0} &=& -\frac{1}{p_{n,m-1}}+\frac{1}{2 \,q}, \nonumber 
\\
 \frac{d u_{n,m}}{d \tau_1}  &=&  
\frac{1}{p_{n,m-1}^2} \left( \frac{1}{ p_{n,m} } \, + 
\frac{1}{p_{n,m-2} }  \right) - \frac{1}{4\, q^3}
 . \nonumber 
\eea

\item It is easy to prove that the two classes of 
symmetries we have constructed commute among themselves, i.e. 
$[\, \widehat X_n^{(k)}, \widehat X_m^{(h)} \, ] = 0$ $\forall \, k,h \in
\mathbb{N}$,
when $\mathbb{D}=0$.
\end{enumerate}

\subsubsection{The non-isospectral hierarchy}

The non-isospectral hierarchy of equations is obtained by looking at those equations which are 
obtained from the Lax equation (\ref{s3a}). Let us denote with $\sigma$ (instead of $\ep$) the continuous group parameter in this case.

The nonlinear differential-difference equations of the non-isospectral hierarchy are given by eq. 
(\ref{lt1}) with 
$
V_n^{(0)} =q_{n}.
$
Notice that $V_n^{(0)}$ is not asymptotically bounded, so that the resulting
difference-differential equations obtained by applying the recursion operator $\mathcal L_n$  are not
bounded in the limit $|n| \rightarrow \infty$. 
However, also in this case we can obtain asymptotically bounded equations by considering the inverse operator 
$\mathcal L_n^{-1}$. 
Taking into account that the solution of the equation 
$\Delta_n^{(-)} \, \alpha_n = \beta$ gives $(\Delta_n^{(-)})^{-1} \beta = \beta\, n + \gamma$,  where $\beta$  
and $\gamma$ are  constants with respect to the variation in $n$ 
(but can depend on $\sigma$ or $m$), we can construct a well defined non-isospectral hierarchy of 
asymptotically bounded equations,
starting from $V_n^{(0)}$  and choosing
$$
f(\l)= 2 \, \l^{-k} \, \left( \frac{\l}{p^2} +1 \right)  \qquad k \in \mathbb{N}.
$$
In this case we get
\beq \label{li2}
\frac{d \eta_{n,m}}{d \sigma_k} = \mathcal L_n^{-k} \,\left( \mathcal L_n^{-1}+ \frac{1}{p^2}\right) \, q_{n,m}.
\eeq

The first equation of the inverse hierarchy (\ref{li2}) is obtained
for $k=0$. The resulting equation is a non-autonomous deformation of eq. (\ref{2}). It reads
\beq \nonumber
\frac{d \eta_{n,m}}{d \sigma_0} = \frac{2 \,n-1}{q_{n-1,m}}- \frac{2 \, n+3}{q_{n+1,m}} + \frac{q_{n,m}}{p^2},
\eeq
which in terms of the field $u_{n,m}$ is given by:
\beq \label{2ab}
\frac{d u_{n,m}}{d \sigma_0} = \frac{2 \, n - u_{n,m} -1}{2 \, p^2} + \frac{ 2 \, n - 1}{q_{n-1,m}}.
\eeq
The higher order equations of the non-isospectral hierarchy (\ref{li2}) are all nonlocal.
In correspondence with eq. (\ref{2ab}) we have the following evolution of the spectral data:
\beq \label{sper}
\frac{d T_m}{d \sigma_0} = 0, \qquad \frac{dR_m}{d \sigma_0} = -\frac{\ri \,\kappa}{p\, (p^2+\kappa^2)} \, R_m,
\eeq
with $d \kappa /d \sigma_0 = \kappa /p^2$.

We can now look for  ($n,m$)-dependent symmetries working at the level of the spectrum, namely by looking at those evolutions of the 
reflection coefficients  in correspondence with equations of the hierarchy (\ref{li2}) which 
 commute with the discrete evolution of the reflection coefficient given by   the lpKdV equation (\ref{kdv}) and which thus satisfy eq. (\ref{li12a})
with $\ep_k$ substituted by $\sigma_k$. If we consider   eq. (\ref{2ab}), as
it  is the only local one 
in the hierarchy (\ref{li2}) it is easy to see that
\beq \nonumber
E_m \, \frac{d R_m}{d \sigma_0} \neq \frac{d R_{m+1}}{d \sigma_0}.
\eeq
To get a symmetry we
could add to eq. (\ref{2ab}) any equation of the isospectral hierarchy multiplied by an arbitrary function of $m$ and $\kappa$, as it
has been done in the case of the 
Toda Lattice \cite{hlrw}.  To do so, let us add an unknown arbitrary function
 $\gamma_m(\kappa;p,q)$ to the evolution (\ref{sper}),
$$
\frac{dR_m}{d \sigma_0} = \left(-\frac{\ri \,\kappa}{p\, (p^2+\kappa^2)} + \gamma_m(\kappa;p,q)\right)\, R_m,
$$
and let us require  that $E_m \, (d R_m /d \sigma_0) = d R_{m+1} / d \sigma_0$
holds. In such a way we find 
\beq
\gamma_m(\kappa;p,q) = -\frac{2 \, \ri \, \kappa \, m}{(q^2+ \kappa^2) \, q}. 
\label{li12qq}
\eeq
It is easy to prove that it is not possible to write the function  $\gamma_m(\kappa;p,q)$
given by eq. (\ref{li12qq}) 
as a finite combination of isospectral evolutions 
 (\ref{li12}). So we are not able to construct in this way non-isospectral symmetries for the lpKdV  equation.

If we are not interested in looking for asymptotic bounded equations we can extract from  the 
non-isospectral hierarchy  (\ref{li2}) the following local equations
\begin{subequations} \label{mn}
\begin{align}
& \frac{d u_{n,m}}{d \omega_0} = u_{n,m} - p\, n \qquad \quad\,
\mbox{for} \qquad \frac{d \lambda}{d\omega_0} = \frac{2}{p^2}, \label{mn1} \\
& \frac{d u_{n,m}}{d \omega_1} = \frac{n}{q_{n-1,m}}   \; \; \qquad  \qquad \mbox{for} 
\qquad \frac{d \lambda}{d\omega_1} = \frac{1}{\lambda}, \label{mn2}
\end{align}
\end{subequations}
where the group parameter is now denoted by $\omega$. Eqs. (\ref{mn}) are not symmetries for the lpKdV equation
(\ref{kdv}). It is possible to show that eq. (\ref{mn2}) is a {\it master symmetry} \cite{fm} 
of eq. (\ref{kdv}). In fact, let us define the
following infinitesimal generators
\bea
\widehat Y_n^{(1)} \doteq \frac{n}{q_{n-1,m}}\, \partial_{u_{n,m}}, \label{yy1} \\ 
\widehat Y_m^{(1)} \doteq \frac{m}{p_{n,m-1}} \,\partial_{u_{n,m}}, \label{yy2}
\eea
where $\widehat Y_m^{(1)}$ is obtained from $\widehat Y_n^{(1)}$
by applying the discrete symmetry $n \leftrightarrow m$, $p \leftrightarrow q$. 
As $\widehat Y^{(1)}_n$ (or equivalently $\widehat Y^{(1)}_m$) is not asymptotically bounded, one cannot associate with the corresponding equation 
(\ref{s1}) the evolution of a reflection coefficient and thus we cannot compute 
the commutation relations $[\, \widehat X^{(k)}_n,  \widehat Y^{(1)}_n\, ]$ for an arbitrary $k$. 
Let us write down the simplest commutation relations:
\beq
\left[\,\widehat X_n^{(0)} , \widehat Y_n^{(1)} \,\right] = -  \widehat X_n^{(1)}, \quad
\left[\,\widehat X_n^{(1)} , \widehat Y_n^{(1)} \,\right] = - 2\,  \widehat X_n^{(2)}, \quad 
\left[\,\widehat X_n^{(2)} , \widehat Y_n^{(1)} \,\right] = - \frac32 \,  \widehat X_n^{(3)}.  \nonumber
\eeq
A similar result could be obtained considering the commutation of $ \widehat X^{(k)}_m$ and  $\widehat Y^{(1)}_m$. As one can see from the above commutation relations  the commutations of $\widehat X^{(k)}_n$  with $Y_n^{(1)}$ and of $\widehat X^{(k)}_m$  with $Y_m^{(1)}$
provide an alternative
constructive tool with respect to the use of the recursive operator to find the 
infinite families of symmetries $\widehat X_n^{(k)},\widehat X_m^{(k)}$ for
the lpKdV equation (\ref{kdv}). 
\subsubsection{Construction of non-autonomous generalized symmetries}

We can now construct some new  ($n,m$)-dependent non-autonomous symmetries of
eq. (\ref{kdv}). 
Such a construction is provided by the following theorem whose proof is straightforward.

\begin{theorem}
Let $\mathbb{D}(u_{n,m},u_{n\pm 1,m},u_{n,m\pm 1},...;p,q) =0$ be an integrable partial difference equation
invariant under the discrete symmetry $n \leftrightarrow m$, $p \leftrightarrow q$.
Let $\widehat Z_n$ be the differential operator
$$
\widehat Z_n \doteq Z_{n} (u_{n,m},u_{n\pm 1,m},u_{n,m\pm 1},...;p,q) \,  \partial_{u_{n,m}},
$$
such that 
\beq
\left. {\rm pr}\, \widehat Z_n \, \mathbb{D}\,  \right|_{\mathbb{D} =0} = 
\alpha \, g_{n,m}(u_{n,m},u_{n\pm 1,m},u_{n,m\pm 1},...;p,q),
\label{pd}
\eeq
where $g_{n,m}(u_{n,m},u_{n\pm 1,m},u_{n,m\pm 1},...;p,q)$ is invariant under 
the discrete symmetry $n \leftrightarrow m$, $p \leftrightarrow q$
and $\alpha$ is an arbitrary constant. Then
\beq
\left. \left( \frac{1}{\alpha}\,{\rm pr}\, \widehat Z_n - 
\frac{1}{\beta}\, {\rm pr}\, \widehat Z_m \right) \, \mathbb{D} \,  \right|_{\mathbb{D} =0} =0,
\nonumber 
\eeq
where the operator 
$
\widehat Z_m \doteq Z_{m} (u_{n,m},u_{n,m\pm 1},u_{n\pm 1,m},...;q,p) \,  \partial_{u_{n,m}}
$
is obtained from $\widehat Z_n$ under $n \leftrightarrow m$, $p \leftrightarrow q$, so that
$$
\left. {\rm pr}\, \widehat Z_m \, \mathbb{D}\,  \right|_{\mathbb{D} =0}= 
\beta \, g_{n,m}(u_{n,m},u_{n\pm 1,m},u_{n,m\pm 1},...;p,q),
$$
being $\beta$ a constant. So 
\beq
\widehat Z_{n,m} \doteq  \frac{1}{\alpha}\,\widehat Z_n - 
\frac{1}{\beta}\, \widehat Z_m \label{se}
\eeq
is a symmetry of $\mathbb{D} =0$.
\end{theorem}

The above theorem provides a constructive tool to obtain generalized symmetries of the form given in
eq. (\ref{pd}) for the lpKdV equation (\ref{kdv}).
Such a situation occurs with the operators $\widehat Y_n^{(1)},\widehat Y_m^{(1)}$ in eqs. (\ref{yy1}--\ref{yy2}) as 
in this case $\alpha \, g_{n,m}=-1$ and $\beta \, g_{n,m}=1$. 
The symmetry $\widehat Y_n^{(1)} + \widehat Y_m^{(1)}$ has been considered by
Tongas in  \cite{to} 
and the invariance condition one gets from it has been derived by Nijhoff and Papageorgiou \cite{np} from a monodromy problem associated with eq. (\ref{lax1}).

A more complicated symmetry for eq. (\ref{kdv})
can be obtained from Theorem 1 by
defining an operator  $\widehat Z_n^{(w)}$, depending parametrically on $w \in
\mathbb{R}$, with
\beq
Z_{n}^{(w)} \doteq \frac{n\, p^w}{q_{n-1,m}} - \frac{p^{w}-q^{w}}{2 \, (p^2-q^2)} 
\left(  \, p\, n - \frac12 \, u_{n,m}\right). \label{ms}
\eeq
In such a case we find:
$$
\alpha \, g_{n,m} =- 
\frac{x_{n,m}^2(p^w -q^w) +4\,p\,q\,x_{n,m}\,(p^{w-1}+q^{w-1})+4\,p^2\,q^2\,(p^{w-2}-q^{w-2})
}{4\, (p+q)\,(x_{n,m}-p+q)},
$$
where $x_{n,m} \doteq u_{n+1,m} -u_{n,m+1}$.
The operator $\widehat Z_n \doteq Z_{n}^{(w)}
\partial_{u_{n,m}}$, with $Z_{n}^{(w)}$ given by
eq. (\ref{ms}), satisfies the condition (\ref{pd}) and thus  the
operator (\ref{se}) 
defines a one-parameter symmetry for the lpKdV
equation (\ref{kdv}). For $w=0$ in eq. (\ref{ms}) we recover the simmetry generated by 
$\widehat Y_n^{(1)} + \widehat Y_m^{(1)}$.
The simmetry obtained by setting $w=1$ in eq. (\ref{ms}) can be also found in \cite{to}.

Let us give here two other examples of functions $Z_{n}^{(w)}$ satisfying condition (\ref{pd}):
\bea
&& Z_{n}^{(w)} \doteq \frac{n\, (p^w+q^w)}{q_{n-1,m}} -
\frac{u_{n,m}}{p^w-q^w}, \label{z1} \\
&& Z_{n}^{(w)} \doteq \frac{n\, (p\,q)^w}{q_{n-1,m}} - \frac{u_{n,m}}{p^w-q^w}, \label{z2}
\eea
with $w \in \RR \setminus \{0\}$.

Using Theorem 1 we can also show that it is not possible to construct a symmetry using eq.  (\ref{mn1}). 
In fact, defining the differential
operators
\beq
\widehat Y_n^{(0)} \doteq (u_{n,m} - p \, n)\, \partial_{u_{n,m}}, \qquad 
\widehat Y_m^{(0)} \doteq  (u_{n,m} - q\, m)\,\partial_{u_{n,m}}, \nonumber
\eeq
we immediately get that condition  (\ref{pd}) is not satisfied as
$$
\alpha \, g_{n,m} = 
\frac{q \, x_{n,m}^2  + 2\, p\,(p-q) \,x_{n,m}- 2\,p^2 (p-q)}{x_{n,m}-p+q},
$$
is not invariant under the exchange   $n \leftrightarrow m$ and $p \leftrightarrow q$.

\subsubsection{Symmetry reduction on the lattice}

Let $\mathbb{D} =0$ be a discrete equation and 
$\widehat X_{n,m} = F_{n,m} ( u_{n,m}, u_{n \pm 1,m}, u_{n, m \pm 1}, \ldots)
\partial_{u_{n,m}}$ one of its symmetry generators. A solution $u_{n,m}$ of $\mathbb{D} =0$
is a {\it  solution invariant} under $\widehat X_{n,m}$ if it satisfies  the equation $\widehat X_{n,m} u_{n,m} =0$.

We now use the notion of invariant solutions of the lpKdV equation
(\ref{kdv}) to construct a particular solution.
Our results are a slight generalization of the
ones contained in  \cite{ns,to}.

Hereafter,  carrying out the  simplifying  transformation
\beq
u_{n,m}\mapsto u_{n,m} \sqrt{p+q} + p\, n +q\, m,\label{map}
\eeq
we shall write the lpKdV equation (\ref{kdv}) as
\beq
(u_{n+1,m} - u_{n,m+1})\,(u_{n+1,m+1} -u_{n,m})= p-q. \label{kdv4}
\eeq

Let us consider the symmetry generator $\widehat X_{n,m} \doteq 
\sqrt{p+q} \, \widehat Z_{n,m}^{(w)} + c \, \widehat X_{n,m}^{1}$, $ c \in \mathbb{R}$, where
$\widehat Z_{n,m}^{(w)} = \widehat Z_{n}^{(w)} + \widehat Z_{m}^{(w)}$, with 
$ Z_{n}^{(w)}$ given in eq. (\ref{ms})
and $\widehat X_{n,m}^{1}$ given in eq. (\ref{ps5a}).
Under the map (\ref{map}), the
differential operator $\widehat X_{n,m}$ reads
\beq
\widehat X_{n,m} = \left( \frac{n \, p^w}{u_{n+1,m}-u_{n-1,m}} + \frac{m \,
  q^w}{u_{n,m+1}-u_{n,m-1}}-
\half \frac{p^w - q^w}{p-q} \, u_{n,m} + c \right)
\partial_{u_{n,m}}. \label{op}
\eeq
Solutions invariant under the symmetry generator (\ref{op}) are solutions of
eq. (\ref{kdv4}) subject to the constraint
\beq
\frac{n \, p^w}{u_{n+1,m}-u_{n-1,m}} + \frac{m \,
  q^w}{u_{n,m+1}-u_{n,m-1}}-
\half \frac{p^w - q^w}{p-q} \, u_{n,m} + c =0. \label{ll}
\eeq
Following \cite{to} we define $a_{n,m} \doteq u_{n+1,m}- u_{n-1,m}$,
$b_{n,m} \doteq u_{n,m+1}- u_{n,m-1}$ and $y_{n,m} \doteq u_{n+1,m+1}-
u_{n,m}$.
Using eq. (\ref{kdv4}) we get
$$
a_{n,m} = y_{n-1,m} + \frac{\delta}{y_{n,m}},
$$
with $\delta \doteq p-q$. On the other hand, we have
\beq
\frac{1}{b_{n+1,m}}= \frac{\delta}{b_{n,m}\, y_{n,m}^2}+\frac{1}{y_{n,m}}.\label{ll2}
\eeq
Eliminating the variable $b_{n,m}$ between eqs. (\ref{ll}--\ref{ll2}) 
we obtain the following second-order difference system:
\begin{subequations} 
\begin{align}
& \frac{p^w \, (n+1)\, \delta}{y_{n}\, y_{n+1}+\delta}+
\frac{p^w \, n\, \delta}{y_{n}\, y_{n-1}+\delta} =
p^w \, (n+1)+ q^w \,m - c \frac{\delta}{y_{n}}+ \nonumber \\
& \qquad +  c \, y_{n} + \frac{p^w-q^w}{2\, \delta} \left(y_{n} \, u_{n} + \delta\,
\frac{u_{n+1}}{y_{n}}\right) ,  \label{pa}\\
& u_{n+1}-u_{n-1} = y_{n-1} +\frac{\delta}{y_n} , \label{paa}
\end{align}
\end{subequations}
where we have dropped the index $m$ in the dependent variables as it enters
just parametrically. Eqs. (\ref{pa}--\ref{paa}) are a system extension of the
alternate discrete Painlev\'e II equation \cite{fok,ns}.

A similar procedure can be performed considering the operators
(\ref{z1}--\ref{z2}). In these cases one obtains, as a reduction
of eq. (\ref{kdv4}), an equation that is equivalent to
the alternate discrete Painlev\'e II equation for any value of $w$.

\section{Concluding remarks} \label{S4}
In this paper we have studied in detail the lpKdV equation and its
symmetries. To do so in a complete way we started from its associated spectral problem. 
The study of the direct and inverse problem allow us to provide the soliton solutions of the nonlinear lattice equation.
As has been shown in \cite{frank,np} the associated spectral problem of the lpKdV equation is given by
the asymmetric discrete Schr\"odinger spectral problem considered by Boiti et
al. \cite{boiti2}. A class of generalized symmetries of the lpKdV equation
is obtained by considering its associated isospectral nonlinear evolution equations. 

It is worthwhile to notice here that the same transformation that exists
between the Volterra equation (\ref{vol}) 
and the discrete KdV equation (\ref{dkdv}) can be written between the
corresponding spectral problems. So part of the
 results we presented could be obtained from the discrete Schr\"odinger
 spectral problem associated with the Volterra 
equation (\ref{vol}). The eq. (\ref{dkdv}) is associated with the spectral problem (\ref{sp}) which, in the variables of the discrete KdV equation, reads
\beq \label{csp}
\psi_{n+2} = \frac{2 \, p}{s_n} \psi_{n+1} + \lambda \, \psi_n,
\eeq
while the spectral problem for eq. (\ref{vol}) is
\beq \label{cspvol}
\phi_{n-1} + a_n \,  \phi_{n+1} = \mu \, \phi_n.
\eeq
By up-shifting eq. (\ref{cspvol}) by one, defining $a_n \doteq s_n \, s_{n-1}$ and considering the trasformation $\phi_n \doteq
f_n(\mu)\, 
g_n(s_n, s_{n \pm 1},...) \, \psi_n$,  we are able to rewrite eq. (\ref{cspvol}) as eq. (\ref{csp}). To do so we  choose  $f_n =
\mu^n$ with $\lambda = - 4 \, p^2 /\mu^2$ and $g_n$ given by the solution of the recursion relation 
$g_{n+1} =g_n / (2\, p \, s_n)$.
From this mapping, but in particular from the relation between $\lambda$ and
$\mu$, we get that the positive 
Volterra hierarchy \cite{hlrw} corresponds to a negative hierarchy in the
asymmetric discrete Schr\"odinger spectral 
problem (\ref{csp}). In this way the Miura transformation between the potential $a_{n,m}$
of the Volterra hierarchy and field 
$u_{n,m}$ of the lpKdV equation is easily obtained  and is given by
$$
a_{n,m} = \frac{4 \, p^2}{(2 \, p - u_{n+2,m} + u_{n,m})(2\, p - u_{n+1,m}+u_{n-1,m})}.
$$

From the two equivalent components of the Lax pair we get two infinite hierarchies of integrable 
symmetries by looking at its isospectral deformations. From the non-isospectral deformations we get an infinite set of master symmetries. 
Apart from the two classes of integrable symmetries, associated with the two equivalent members of the Lax pair,  
we have obtained an infinite set of symmetries starting from the master symmetries. 
It is an open problem to 
prove whether the symmetries $\widehat Z^{(w)}$ are integrable, possibly by providing their Lax
pair and to compute their commutation relations. In any case we can use them to provide solutions by symmetry reducing the lpKdV.

In conclusion we have shown that the lpKdV equation has a lot of
symmetries. It is left to future work the proof if all are independent or if,
the two sequences associated with the spectral problems in $n$ and $m$
respectively are dependent under the equation. 

It is also to be clarified the
role of the $\widehat Z^{(w)}$ symmetries, related to the master symmetry.
It is an open problem to see whether the other discrete symmetries
might give rise to new continuous symmetries. We leave to a future work the use of the $\widehat Z^{(w)}$ symmetries here obtained to check the integrability
of the (local) dNLS equations derived in \cite{JLP, levi, levipetrera}. 

\vspace{.3truecm}

After we finished this article a referee for our recent paper \cite{lps}
pointed out to us the article by
Rasin and Hydon \cite{rh} submitted to {\it Stud. Appl. Math.} and
available in the personal web page of the authors. 
In this article the authors construct the symmetries for all quad-graph equations contained in the Adler-Bobenko-Suris list \cite{abs}.
Let us notice here that in the present paper we have just considered the $H_1$ equation
but with much more details. In particular we already answer in this case
to some of the questions in their conclusions. Moreover we have obtained five point symmetries that they have not been able to find.

\section*{Acknowledgements}
The authors were partially supported by  PRIN Project ``SINTESI-2004'' of the
Italian Minister for  
Education and Scientific Research. This research is part of a joint Italian
Russian research project 
``Classification of integrable discrete and continuous
models" financed by Consortium EINSTEIN and Russian Foundation for Basic
Research. The authors thank C. Scimiterna for some fruitful discussions.


\end{document}